\documentclass[preprint,12pt,authoryear]{elsarticle}

\usepackage{amssymb}
\usepackage{amsmath}
\usepackage{url}
\usepackage{lineno}

\usepackage{float}

\usepackage[colorlinks=true,linkcolor=blue,citecolor=blue,urlcolor=blue]{hyperref}

\journal{Nuclear Physics B}

\begin{document}

\begin{frontmatter}



\title{Consistency between dynamical modeling and photometrically derived masses of fireballs} 

\author[label1]{Eloy Peña-Asensio}
\author[label2,label3,label4]{Maria Gritsevich}

\affiliation[label1]{
    organization={Department of Applied Mathematics and Aerospace Engineering, Universitat d'Alacant},
    addressline={P.O. Box 99},
    city={Alacant},
    postcode={03690},
    country={Spain}
}

\affiliation[label2]{
    organization={Faculty of Science, University of Helsinki, Gustaf H\"{a}llstr\"{o}min katu 2},
    city={Helsinki},
    postcode={FI-00014},
    country={Finland}
}

\affiliation[label3]{
    organization={Instituto de Astrofisica de Andalucia (IAA-CSIC)},
    addressline={Glorieta de la Astronomia s/n},
    city={Granada},
    postcode={E-18008},
    country={Spain}
}

\affiliation[label4]{
    organization={Institute of Physics and Technology, Ural Federal University},
    addressline={Mira str. 19},
    city={Ekaterinburg},
    postcode={620002},
    country={Russia}
}

\cortext[cor1]{Corresponding author: eloypa@ua.es}

\begin{abstract}
We present a three-point inverse solution for reconstructing meteoroid deceleration and mass-loss histories from sparse observations constrained only by the entry, peak-brightness, and terminal points. The method combines the $\alpha$–$\beta$ analytical formalism with a derivative-free global optimizer and a numerical inversion of the height–velocity relation, enabling the retrieval of physically consistent solutions even when full velocity profiles are unavailable. Applied to the 2017–2018 European Fireball Network (EN) catalog, the approach achieves an 88\% convergence rate when fitting only height–velocity pairs, and 63\% when terminal and initial masses are also imposed. 52\% of mass-constrained solutions (34\% overall) yield bulk densities consistent with their $PE$ classes, with higher strength emerging as the primary discriminator among events retaining coherent classifications when only 3 points are used as input data. Rapidly evolving high-energy, high-mass events show the largest incompatibility with the $\alpha$–$\beta$ model. The inversion produces a continuous bulk-density distribution spanning $\sim$300–4000 kg\,m$^{-3}$, in contrast to the discrete densities fixed by $PE$-based categories. The EN fireball dataset is now supplemented with self-consistent $\alpha$ and $\beta$ estimates. 
\end{abstract}


\begin{highlights}
\item A three-point inversion method reconstructs meteoroid deceleration and mass loss from sparse datasets.
\item Application to the EN 2017–2018 catalog yields self-consistent $\alpha$ and $\beta$ values for 88\% of events.
\item 34\% of EN fireballs yield $\alpha$-$\beta$ solutions consistent with photometrically derived masses.
\item High radiated energy and large initial mass primarily control inconsistencies between dynamical and photometric estimates.
\end{highlights}


\end{frontmatter}




\section{Introduction} \label{sec:intro}

Quantifying the aerodynamic and ablative evolution of meteoroids during atmospheric entry remains central to understanding the physical diversity of small Solar System bodies and their delivery pathways to Earth. Meteoroid atmospheric entries producing fireball observations \citep{Ceplecha1998SSRv, Koschny2017JIMO, Silber2018AdSpR} encode this evolution through their radiometric and dynamical signatures, yet most global fireball networks lack the continuous velocity and brightness records required to fully resolve deceleration and mass loss. As a result, modeling efforts used to rely on single-body ablation modeling constrained by sparse measurements or on population-level proxies, such as the PE classification \citep{Ceplecha1976JGR}, to infer meteoroid structure from gross entry parameters. While these approaches offer useful taxonomies, they provide limited insight into the continuous distributions of bulk density, strength, and mass-loss behavior that likely characterize real meteoroids across size scales \citep{moreno2020physically}.


The long-standing mismatch between dynamical and photometric mass estimates can be mitigated when appropriate bulk densities were enforced or fragmentation was explicitly introduced into the modeling \citep{Ceplecha1998SSRv, Ceplecha2005, Popova2019msmebook, Borovicka2019msmebook, Gritsevich2008Validity}. Other approaches combine astrometric, radiometric, and fragmentation information in joint inversions, retrieving deceleration and mass-loss histories within a unified scheme \citep{Borovicka1998AA334713B, Borovicka2020AJ, VIDA2024115842, McMullan2024MPS59927M, Egal2025NatAs91624E}. Although such methods increase the dimensionality of the parameter space, they improve consistency between the dynamical and luminous constraints.


Recent work has extended these efforts through inverse and optimization-based approaches to meteoroid characterization \citep{Tarano2019Icar, VIDA2024115842}. However, such methods require high-fidelity velocity time series or event-specific modeling, limiting their applicability to sparse datasets. Consequently, vast portions of archival fireball data remain effectively underexploited: many events report only three physical constraints---the entry point, the altitude of peak brightness, and the terminal point without error constrains---insufficient for traditional inverse modeling. A scalable inversion framework capable of extracting physically consistent dynamical histories from such minimal information is therefore essential, both for reanalyzing existing catalogs and for enabling future network interoperability.

In our previous study \citep{EloyMaria2025}, we delimited the fireball parameter space using the height, velocity, and mass at the entry and terminal points using the purely dynamical $\alpha$-$\beta$ approach, where $\alpha$ is the ballistic coefficient and $\beta$ is the mass-loss parameter. Here we extend that strategy: each event is now fitted with the same entry and terminal constraints plus the height–velocity pair at peak brightness. By means of using a general three-point inversion methodology that reconstructs meteoroid deceleration and mass-loss histories, this extension serves three purposes: (i) it demonstrates that unique and physically consistent solutions can be recovered even when only three trajectory points are available, a common limitation in several widely used meteor databases (e.g., GMN, CAMS, FRIPON, SonotaCo), where only summary trajectory parameters are available; (ii) it provides refined physical parameters for several hundred additional fireballs, strengthening the statistical basis for meteoroid research; (iii) it constrains the conditions under which purely dynamical and photometrically supported models remain consistent. Application to the 2017–2018 European Fireball Network (EN) fireball catalog \citep{Borovicka2022AA_I, Borovicka2022AA_II} demonstrates both the robustness and the limits of the $\alpha$-$\beta$ description across a broad range of entry conditions. The resulting database supports automated fireball classification, guides rapid meteorite-recovery efforts, and strengthens tools for monitoring and mitigating near-Earth object (NEO) hazards.

This paper is organized as follows. Section \ref{sec:model_class} summarizes the dynamical meteoroid entry model and the parameterization adopted for trajectory and mass-loss calculations. Section \ref{sec:inferring} details the inversion framework used to infer physical properties from sparse observational constraints. Section \ref{sec:stats} presents the statistical framework used for the comparison of sub-populations. Section \ref{sec:results} presents the derived parameters for the expanded dataset and discusses the implications for fireball classification and meteoroid characterization. Section \ref{sec:conclusions} outlines the main findings and prospects for future applications.

\section{Dynamical meteoroid entry model} \label{sec:model_class}

\subsection{Motion equations}

The meteoroid’s dynamics can be written as (adapted from \citet{gritsevich2017constraining}):

\begin{equation}
M\frac{dV}{dt}=-\frac{1}{2}c_{d}\rho V^{2}S, \label{1}
\end{equation}

\begin{equation}
\frac{dh}{dt}=-V\sin \gamma, \label{2}
\end{equation}

\begin{equation}
H^{\ast }\frac{dM}{dt}=-\frac{1}{2}c_{h}\rho V^{3}S. \label{3}
\end{equation}

These equations describe meteoroid motion in planetary atmospheres \citep{CHRISTOU2024116116} and depend on the physical quantities defined in \ref{sec:defs}. Many solution strategies exist; most require assumptions about unmeasured parameters. We instead recast the system in nondimensional form via dimensional analysis because this reduces the number of governing scales and clarifies parameter dependencies \citep{Stulov1997ApMRv, Gritsevich2006SoSyR}.

Following the Buckingham $\pi$ theorem (see Appendix B of \citet{EloyMaria2025}), we construct independent dimensionless groups. Eliminating explicit time with Eq.~\ref{2} leaves $M$, $V$, and $h$ as independent variables; atmospheric density $\rho(h)$ and cross-sectional area $S$ (expressible through $M$) act as dependent functions. Each dimensional variable is scaled by an appropriate characteristic quantity (definitions in \ref{sec:defs}).

In these dimensionless variables, the coupled mass- and velocity-loss equations as functions of height read \citep{Gritsevich2007SoSyR}:

\begin{equation}
m\frac{dv}{dy}=\frac{1}{2}c_{d}\frac{\rho_{sl}h_0S_{beg}}{M_{beg}}\frac{\rho vs}{\sin \gamma }, \label{4a}
\end{equation}

\begin{equation}
\quad \frac{dm}{dy}=\frac{1}{2}c_{h}\frac{\rho_{sl}h_0S_{beg}}{M_{beg}}\frac{V_{beg}^{2}}{H^{\ast }}\frac{\rho v^2s}{\sin \gamma }. \label{4b}
\end{equation}

Assuming $s = m^{\mu}$, an isothermal atmosphere described by $\rho = e^{-y}$ (see \cite{Lyytinen2016PSS} for alternative implementations with other atmospheric models), and applying the boundary conditions $y \to \infty$, $v = 1$, and $m = 1$, the analytical solution of Eqs.~(\ref{4a}) and (\ref{4b}) can be expressed as:

\begin{equation}
m(\beta,\mu,v)= e^{-\beta \frac{1-v^{2}}{1-\mu }}\label{6_mass}
\end{equation}

and

\begin{equation}
y(\alpha,\beta,v)=\ln (2\alpha) +\beta -\ln (\overline{E}i(\beta )-\overline{E}i(\beta v^{2})), \label{eq_y}
\end{equation}

where 
\[
\overline{Ei}(x)=\int_{-\infty }^{x}\frac{e^{z}dz}{z} 
\]

is the exponential integral (see Appendix C of \citet{EloyMaria2025}), where the ballistic coefficient $\alpha$ and the mass-loss parameter $\beta$ are defined as:

\begin{equation}
\alpha =\frac{c_{d}\rho_{sl}h_0S_{beg}}{2M_{beg}\sin \gamma }, \label{a}
\end{equation}

\begin{equation}
\beta =\frac{(1-\mu )c_{h}V_{beg}^{2}}{2c_{d}H^{\ast }}. \label{b}
\end{equation}

This approach simplifies fireball flight modeling by reducing all unknowns to two dimensionless parameters, $\alpha$ and $\beta$, which have clear physical meanings and can be uniquely derived from observations for each event. In practice, fireballs are classified into comparable outcome groups based on their specific $\alpha$ and $\beta$ values \citep{Gritsevichcite2009Classification, Gritsevich2012CosRe}. The ballistic coefficient $\alpha$ represents the atmospheric mass encountered along the trajectory, proportional to the column mass intercepted by the entry cross-section $S_{beg}$ and scaled by the meteoroid’s pre-atmospheric mass. Physically, $\alpha$ reflects the altitude and intensity of aerodynamic drag experienced during flight. The mass-loss parameter $\beta$ characterizes the rate of ablation, interpreted as the ratio between the meteoroid’s initial kinetic energy at atmospheric entry and the energy required to completely disintegrate it during passage through the atmosphere.

When analyzing fireball data, the $\alpha$–$\beta$ approach provides a compact and physically meaningful description of meteoroid atmospheric entry \citep{Gritsevich2007SoSyR, Gritsevich2008SoSyR}. It has been widely applied to interpret variations in deceleration rate, deceleration height, and ablation efficiency among observed events. Within this parameter space, meteoroids that penetrate more deeply and experience lower rates of mass loss generally occupy regions associated with a higher probability of meteorite survival, typically defined in terms of a minimum terminal mass required to exceed the meteorite-producing threshold. However, the transition between survival and complete ablation is continuous rather than sharply bounded, reflecting the gradual nature of mass loss and the absence of strict thresholds.

The $\alpha$ and $\beta$ parameters have also been extensively used to estimate terminal heights \citep{Gritsevich2008DokPh, moreno2015new, Moreno2017ASSP,  Sansom2019ApJ, EloyMaria2025} and to investigate light-curve behaviour, including the effects of fragmentation, rotation, and luminous efficiency \citep{Gritsevich2011Icar, Bouquet2014PSS, Drolshagen2021AAa, Drolshagen2021AAb}. These applications highlight the utility of the formalism as a physically interpretable, low-dimensional framework for describing meteoroid atmospheric entry, while also emphasizing its limitations when single, abrupt fragmentation events rather than continuous mass loss characterize the dynamics.

\subsection{Estimation of initial and terminal mass}

When interpreting meteor phenomena, dynamic parameters generally refer to the main leading fragment unless explicitly stated otherwise or unless individual fragment trajectories are analyzed separately \citep{Moilanen2021LPICo26096288M}. Determining the mass evolution of this primary body along the luminous trajectory requires knowledge of $\beta$ (or $\sigma$) and the shape-change coefficient $\mu$, which typically ranges from 0 (no rotation) to 2/3 (isotropic surface ablation) \citep{Gritsevich2006SoSyR, Bouquet2014PSS, Sansom2019ApJ}. In the classical inverse-problem approach, all mass-loss processes, including fragmentation, are incorporated into the model, with the main fragment mass decreasing exponentially as described by Eq.~\ref{6_mass}. Bayesian filtering and Monte Carlo simulations can also be used to estimate both the mass of the main body and individual fragments \citep{Sansom2017, Moilanen2021MNRAS, Gritsevich2024}. These estimates, however, remain sensitive to assumptions regarding the atmospheric model \citep{Lyytinen2016PSS}.

Traditionally, photometric methods estimate meteoroid mass by evaluating the fraction of kinetic energy converted into light during the luminous flight \citep{McCrosky1968}. This approach typically assumes constant velocity and relies on uncertain luminous efficiency values, making it less reliable, particularly for meteorite-dropping fireballs \citep{Ceplecha1996physics, Gritsevich2008Validity, Gritsevich2011Icar}.

Following the dynamic approach of \citet{Gritsevich2009AdSpR}, the initial mass of a meteoroid at the start of its luminous trajectory can be expressed from Eq.~\ref{a} as:

\begin{equation}
M_{beg}=\left(\frac{1}{2} \frac{c_d A_{beg} \rho_{sl} h_0}{\alpha \rho_m^{2/3} \sin\gamma}\right)^{3}.
	\label{eq:mass_init}
\end{equation}

Previous research has demonstrated that pre-atmospheric mass estimates derived through this method show good agreement with pre-atmospheric sizes inferred from laboratory analyses of recovered meteorites based on cosmogenic radionuclide measurements \citep{gritsevich2008estimating, gritsevich2017constraining, kohout2017annama, meier2017park, Gritsevich2024}, despite the fact that the method primarily reflects the deceleration of the main fragment and does not explicitly model fragmentation.

The terminal mass can be determined by substituting the final observed velocity and the calculated pre-atmospheric mass into Eq.~\ref{6_mass}:

\begin{equation}
M_{ter}=M_{beg} e^{-\frac{\beta}{1-\mu} \left(1-\left(\frac{V_{ter}}{V_{beg}}\right)^2 \right)}.
	\label{eq:mass_terminal}
\end{equation}

This estimate can be refined by accounting for the additional minor ablation occurring below the terminal point of the fireball \citep{Moilanen2021MNRAS}.

\subsection{Classification of fireballs}

Classifying fireballs is crucial for characterizing the physical properties, origin, and atmospheric behavior of meteoroids, as well as for rapidly evaluating their potential impact outcomes. A widely used classification method is the $PE$ criterion proposed by \citet{Ceplecha1976JGR}, which quantifies the penetration capability of meteoroids. This empirical relation, derived from total photometric data, is used to distinguish different meteoroid types and is calculated as:

\begin{equation}
P_E = \log \rho_{ter} - 0.42 \log M_{phot} + 1.49 \log V_{beg} - 1.29 \log \cos z,
\label{eq:PE}
\end{equation}

where $\rho_{ter}$ is the atmospheric density at the terminal height of the fireball (in $g\,cm^{-3}$), $M_{phot}$ is the photometric mass in grams calculated using the original luminous efficiency $\tau$ \citep{Ceplecha1976JGR}, $V_{beg}$ is the entry velocity in $km\,s^{-1}$, and $z$ is the zenith distance of the apparent radiant. Notably, the $PE$ criterion does not rely on assumptions about the ablation coefficient or the meteoroid bulk density. Instead, it uses directly observable parameters, including atmospheric density at the terminal point, entry velocity, trajectory slope, and light-curve-derived mass. However, for the criterion to be applied, both the initial and terminal trajectory points must be recorded.

McCrosky and Ceplecha classified 232 Prairie Network fireballs into four groups according to their $PE$ values. They associated each group with representative bulk densities and ablation coefficients, providing a practical interpretation of meteoroid behavior during atmospheric entry. The four fireball classes (I, II, IIIA, and IIIB, ordered from strongest to weakest) are defined by boundary $PE$ values of –4.60, –5.25, and –5.70. Once the $PE$ value is determined, any event can be assigned to one of these groups, and default bulk density and ablation coefficient values are applied accordingly.

\citet{moreno2020physically} proposed an alternative classification based on the $\alpha$ and $\beta$ parameters, which provides a physically consistent counterpart to the $PE$ criterion. Using analytical derivations and Prairie Meteor Network observations, they showed that $y_{ter} = \ln(2 \alpha \beta)$ ensures continuity in the derived ablation coefficients and bulk densities. Unlike the $PE$ method, this approach does not require assumptions about luminous efficiency or photometric mass (as in Eq.~\ref{eq:PE}), making it more robust and flexible. It avoids sharp group boundaries and eliminates the need for empirical default parameters such as preset bulk densities and ablation coefficients, offering a continuous and physically grounded framework better suited for characterizing meteoroid properties and atmospheric behavior. This further highlights the advantage of reducing the number of unknowns, as noted at the beginning of this section.

More recently, by examining the dependence of fragmentation behavior on mass and velocity, \citet{Borovicka2022AA_II} introduced a new parameter to evaluate material strength: the pressure resistance factor, or pressure factor ($Pf$). This parameter, defined as the maximum dynamic pressure experienced by large meteoroids, is expressed as:

\begin{equation}
Pf = 100  \rho_{max}  \cos z^{\ -1}  M_{phot}^{\ -1/3}  V_{beg}^{\ -3/2},
\label{eq:Pf}
\end{equation}

where $\rho_{max}$ represents the maximum dynamic pressure in $MPa$. Five strength categories, $Pf$-I to $Pf$-V (from strongest to weakest), are defined with the following boundaries: $Pf > 0.85$ for $Pf$-I, $0.27 < Pf \leq 0.85$ for $Pf$-II, $0.085 < Pf \leq 0.27$ for $Pf$-III, $0.027 < Pf \leq 0.085$ for $Pf$-IV, and $Pf \leq 0.027$ for $Pf$-V.

Fireballs producing meteorites are generally classified as $PE$ type I or $Pf$-I. Alternatively, the $\alpha$–$\beta$ framework, combined with terminal mass calculations, provides a visual classification into three regions: ``Likely fall,'' ``Possible fall,'' and ``Unlikely fall'' \citep{Sansom2019ApJ, moreno2020physically, Boaca2022ApJ, EloyMaria2025}. Unlike the $PE$ and $Pf$ criteria, these regions are not defined by fixed thresholds in $\alpha$ and $\beta$, but by continuous combinations of these parameters that yield terminal masses above or below a given survival threshold. A terminal mass of 0.05 kg is commonly adopted as a practical threshold for meteorite recovery \citep{halliday1996detailed}, although this value may be adjusted based on practical considerations depending on complexity of the terrain for recovering meteorites. The ``Possible fall'' region is bounded analytically, with the upper limit corresponding to non-rotating meteoroids ($\mu = 0$) and the lower limit to the maximum rotation case ($\mu = 2/3$) \citep{Sansom2019ApJ}.

The broader application of $\alpha$ and $\beta$ for classifying meteoroid and asteroid atmospheric entry outcomes was first proposed by \citet{Gritsevichcite2009Classification, gritsevich2011DokPh}. Depending on these parameters, possible outcomes range from the formation of a single large crater, to fragmentation with multiple crater-forming impacts, to ablation-dominated entry producing small surviving fragments, or complete atmospheric disintegration and vaporization \citep{Gritsevich2012CosRe, Turchak2014JTAM}.

\section{Determining meteoroid parameters} \label{sec:inferring}

The atmospheric flight characterization framework outlined above highlights the importance of measuring velocity as a function of height to properly describe meteoroid entry dynamics. In this work, however, we aim to reconstruct velocity profiles and mass-loss rates without requiring full observational datasets, allowing analysis when observations are incomplete or unavailable. The inverse problem is formulated as follows: given three observational points along the trajectory—the initial, peak-brightness, and terminal points—each defined by height, velocity, and mass at the initial and terminal point, we seek to determine the set of physical parameters ($\alpha$ and $\beta$, and $\rho_m$ and $\mu$ when considering masses) that best reproduces these constraints within the ($\alpha$-$\beta$)  model. To assess the impact of including mass constraints, we perform the inversion both with and without enforcing mass consistency, allowing a direct comparison of the resulting solutions. In contrast to our previous study \citep{EloyMaria2025}, where only the entry and terminal points were used and the solution was constrained to regions of the parameter space, the inclusion of the peak-brightness point reduces degeneracy and allows the retrieval of specific solutions. This shifts the problem from assessing potential compatibility to enforcing explicit agreement between models and observations. The primary outputs of the inversion are the parameters $(\alpha, \beta, \rho_m, \mu)$, from which the full velocity profile and mass evolution are reconstructed.


As a case study, we use the EN catalog comprising 824 fireballs observed between 2017 and 2018 \citep{Borovicka2022AA_I, Borovicka2022AA_II}. Although this dataset includes a large number of events with computed physical parameters, it lacks raw measurements and derived quantities such as complete velocity profiles. Nonetheless, it provides initial, peak-brightness, and terminal values of key luminous-phase parameters, as well as integrated energy estimates and classifications using $PE$ and $Pf$. The inputs are taken directly from the EN catalog and correspond to the reported parameters ``H-beg'', ``H-max'', and ``H-ter'' (initial, peak-brightness, and terminal heights), ``Vinf'', ``Vmax'', and ``Vter'' (initial, peak-brightness, and terminal velocities), as well as the photometric mass (``Mass'') and terminal dynamic mass (``TerMass''). It is important to note that, in this catalog, $M_{beg} = M_{pho}$, whereas $M_{ter}$ was dynamically computed assuming a bulk density of $\rho_m = 3000 \, kg\,m^{-3}$ and $(1/2) c_d A = 0.7$ \citep{Borovicka2022AA_I}.

Prior to the inversion, a data-quality filtering step is applied to ensure internal consistency of the input parameters. Events are excluded if they violate the expected monotonic behavior of the trajectory, namely $h_{beg} > h_{bri} > h_{ter}$ and $V_{beg} > V_{bri} > V_{ter}$, or if cross-inconsistencies between height and velocity evolution are present. Additionally, events with missing or null values in the peak-brightness parameters ($h_{bri}$ or $V_{bri}$) are removed. After applying these criteria, 809 out of the initial 824 events are retained for analysis.

In \citet{EloyMaria2025}, we inferred fireball parameters using the Differential Evolution (DE) algorithm, a population-based stochastic global optimization method \citep{Storn1997JGOpt, Wormington1999RSPTA, Qiang2014}. In this work, the parameter inference is performed using the Covariance Matrix Adaptation Evolution Strategy (CMA-ES), a stochastic, derivative-free algorithm designed for continuous, non-linear, and multimodal optimization problems. Although the transition to CMA-ES was expected to provide a significant improvement in convergence and solution quality, we find that both methods yield comparable performance for the present problem. CMA-ES iteratively samples candidate solutions from a multivariate normal distribution, whose mean, covariance matrix, and step size are continuously adapted based on the best-performing solutions at each iteration. This adaptive mechanism allows efficient exploration of complex parameter spaces without requiring gradient information, making it well suited for the present inverse problem. We use the Python implementation provided by the \texttt{cma} package \citep{hansen2019pycma}, publicly available on PyPI.

To constrain the parameter space for the EN catalog, we adopt the ranges summarized in \citet{Ceplecha1998SSRv} (Table XVII). For meteoroids without specific compositional classification, the bulk density is restricted to $250 \leq \rho_m \leq 4000 \, \mathrm{kg\,m^{-3}}$, scaled by a factor $0.95 \leq n \leq 1.05$ to account for uncertainties. For the single event explicitly classified as iron \citep{Borovicka2022AA_II}, the density range is narrowed to $6000 \leq \rho_m \leq 8000 \, \mathrm{kg\,m^{-3}}$, applying the same scaling factor. The optimization is performed within the bounds $\mu \in [0, 2/3]$, $\alpha \in [0.1, 10^{4}]$, $\beta \in [0.01, 250]$, and $\rho_m \in [\rho_{m,\mathrm{inf}}, \rho_{m,\mathrm{sup}}]$. The drag component $c_d A$ is modeled as a stochastic variable sampled from a normal distribution, $c_d A \sim \mathcal{N}(1.7,\,0.3)$ \citep{Gritsevich2008SoSyR, Gritsevich2024}. These constraints are applied consistently across the catalog, with the density range adjusted accordingly for the iron-classified event. The optimization is then performed over the parameter vector $(\alpha, \beta, \rho_m, \mu)$; or $(\alpha, \beta)$ when not considering masses, whose values are adjusted to minimize the residuals defined below. The search is initialized from representative values within the prescribed bounds (taken at the midpoint of each interval).

To ensure physically consistent solutions, only optimization results meeting specific convergence criteria were accepted. The initial mass had to be within a factor of five of the value reported in the catalog. This follows the findings of \citet{Lyytinen2016PSS}, who showed that when using a realistic atmospheric model instead of a simple exponential one, the initial mass can vary by up to a factor of five. The terminal mass had to be within a factor of two of the catalog value, with an additional absolute tolerance of one gram to allow for very small deviations.

The terminal height and the height at the brightness peak were required to match the observed values within one hundred meters. The velocities at the initial, brightness, and terminal points had to be very close to the catalog values: the initial velocity was allowed to differ by up to 0.75 kilometers per second, while the velocities at the brightness peak and at the terminal point could not deviate by more than 0.1 kilometers per second. The more permissive tolerance for the initial velocity is justified by \citet{Vida2018MNRAS}, who showed that meteoroids can decelerate by as much as 0.75 kilometers per second before the luminous phase begins, and also because the initial velocities in the EN catalog represent the hypothetical value the meteoroid would have had without any atmospheric influence.

To reduce the risk of the optimizer becoming trapped in local minima, each event is optimized 100 times, with a maximum of 10,000 iterations allowed per run.

Based on these considerations, we formulate the inverse problem under two complementary configurations, both relying on the same three observational points (initial, peak-brightness, and terminal). In the first case, a mass-constrained inversion is performed, in which the parameters $(\alpha, \beta, \rho_m, \mu)$ are inferred by fitting both the kinematic and mass-related constraints. In the second case, only height--velocity constraints are considered, where the free parameters reduce to $(\alpha, \beta)$. These two formulations are presented separately in the following subsections, where the corresponding objective functions and constraints are defined explicitly. This distinction allows us to assess the impact of including mass constraints on the inferred solutions.

\subsection{Three-point inversion with mass constraints}

In this first case, we define the objective function to be minimized as:

\begin{equation}
\begin{aligned}
    Q : \mathbb{R}^4 &\rightarrow \mathbb{R}\\
    Q(\alpha, \beta, \rho_m, \mu) &= \mathfrak{r}(h_{beg}, V_{beg}, M_{beg}, h_{bri}, V_{bri}, h_{ter}, V_{ter}, M_{ter}),
\end{aligned}
\label{eq:my_equation_three_point}
\end{equation}

where $\mathfrak{r}^{(t)}$ is the sum of absolute residuals at iteration $t$, now extended to include the brightness point. The individual residuals are:

\begin{equation}
    \mathfrak{r}_{v_{beg}}^{(t)} =
    y^{-1}_{beg}\!\left(v_{beg};\,\alpha^{(t)},\beta^{(t)}\right) - 1,
\end{equation}

\begin{equation}
    \mathfrak{r}_{M_{beg}}^{(t)} =
    \left(\frac{1}{2}\frac{\rho_{sl} h_0}
    {\alpha^{(t)}\left(\rho_m^{(t)}\right)^{2/3}\sin\gamma}\right)^{3}
    - M_{beg},
\end{equation}

\begin{equation}
    \mathfrak{r}_{y_{bri}}^{(t)} =
    \ln 2\alpha^{(t)} + \beta^{(t)}
    - \ln\!\left(\overline{E}i(\beta^{(t)}) -
    \overline{E}i\!\left(\beta^{(t)}
    \left(\frac{V_{bri}}{V_{beg}}\right)^2\right)\right)
    - \frac{h_{bri}}{h_{0}},
\end{equation}

\begin{equation}
    \mathfrak{r}_{y_{ter}}^{(t)} =
    \ln 2\alpha^{(t)} + \beta^{(t)}
    - \ln\!\left(\overline{E}i(\beta^{(t)}) -
    \overline{E}i\!\left(\beta^{(t)}
    \left(\frac{V_{ter}}{V_{beg}}\right)^2\right)\right)
    - \frac{h_{ter}}{h_{0}},
\end{equation}

\begin{equation}
    \mathfrak{r}_{M_{ter}}^{(t)} =
    M_{beg}^{(t)}\,
    \exp\!\left[-\frac{\beta^{(t)}}{1-\mu^{(t)}}
    \left(1-\left(\frac{V_{ter}}{V_{beg}}\right)^2\right)\right]
    - M_{ter}.
\end{equation}

The total residual to be minimized is:

\begin{equation}
    \mathfrak{r}^{(t)} =
    c_1\left|\mathfrak{r}_{v_{beg}}^{(t)}\right| +
    c_2\left|\mathfrak{r}_{M_{beg}}^{(t)}\right| +
    c_3\left|\mathfrak{r}_{y_{bri}}^{(t)}\right| +
    c_4\left|\mathfrak{r}_{y_{ter}}^{(t)}\right| +
    c_5\left|\mathfrak{r}_{M_{ter}}^{(t)}\right|,
\end{equation}

where the constants $c_i$ weight the residuals according to the desired fitting accuracy. Heights and velocities are mutually dependent, leading to implicit equations; initial velocity is solved numerically, while initial height is not explicitly resolved. Conversely, the terminal and brightness heights are determined analytically, whereas their associated velocities are indirectly computed.

The physical constraints applied during the minimization are:

\begin{equation}
\begin{aligned}
& \underset{(\alpha, \beta, \rho_m, \mu) \in \mathbb{R}^4}{\text{minimize}}
& & Q(\alpha, \beta, \rho_m, \mu), \\
& \text{subject to}
& &
\begin{cases}
\alpha > 0, \\
\beta > 0, \\
\rho_{m,l} \leq \rho_m \leq \rho_{m,u}, \\
0 \leq \mu \leq 2/3.
\end{cases}
\end{aligned}
\end{equation}

\subsection{Three-point inversion without mass constraints}

In this kinematic formulation, only the height--velocity constraints at the three observational points (initial, peak-brightness, and terminal) are enforced. In this case, the mass equations are not evaluated, and the free parameters reduce to $(\alpha, \beta)$.

The objective function to be minimized is defined as:

\begin{equation}
\begin{aligned}
    Q : \mathbb{R}^2 &\rightarrow \mathbb{R}\\
    Q(\alpha, \beta) &= \mathfrak{r}(h_{beg}, V_{beg}, h_{bri}, V_{bri}, h_{ter}, V_{ter}),
\end{aligned}
\end{equation}

where $\mathfrak{r}^{(t)}$ is the sum of absolute residuals at iteration $t$. The individual residuals are:

\begin{equation}
    \mathfrak{r}_{v_{beg}}^{(t)} =
    y^{-1}_{beg}\!\left(v_{beg};\,\alpha^{(t)},\beta^{(t)}\right) - 1,
\end{equation}

\begin{equation}
    \mathfrak{r}_{y_{bri}}^{(t)} =
    \ln 2\alpha^{(t)} + \beta^{(t)}
    - \ln\!\left(\overline{E}i(\beta^{(t)}) -
    \overline{E}i\!\left(\beta^{(t)}
    \left(\frac{V_{bri}}{V_{beg}}\right)^2\right)\right)
    - \frac{h_{bri}}{h_{0}},
\end{equation}

\begin{equation}
    \mathfrak{r}_{y_{ter}}^{(t)} =
    \ln 2\alpha^{(t)} + \beta^{(t)}
    - \ln\!\left(\overline{E}i(\beta^{(t)}) -
    \overline{E}i\!\left(\beta^{(t)}
    \left(\frac{V_{ter}}{V_{beg}}\right)^2\right)\right)
    - \frac{h_{ter}}{h_{0}}.
\end{equation}

The total residual to be minimized is:

\begin{equation}
    \mathfrak{r}^{(t)} =
    c_1\left|\mathfrak{r}_{v_{beg}}^{(t)}\right| +
    c_2\left|\mathfrak{r}_{y_{bri}}^{(t)}\right| +
    c_3\left|\mathfrak{r}_{y_{ter}}^{(t)}\right|.
\end{equation}

The physical constraints applied during the minimization are:

\begin{equation}
\begin{aligned}
& \underset{(\alpha, \beta) \in \mathbb{R}^2}{\text{minimize}}
& & Q(\alpha, \beta), \\
& \text{subject to}
& &
\begin{cases}
\alpha > 0, \\
\beta > 0.
\end{cases}
\end{aligned}
\end{equation}

This formulation provides a reference solution independent of mass assumptions and is used to quantify the impact of including mass constraints in the inversion.

For both fitting schemes, to compute the residuals in each iteration, the heights are computed directly using Eq.~\ref{eq_y}, but the velocity cannot be obtained analytically because the exponential integral has no inverse closed-form solution. Thus, to invert Eq.~\ref{eq_y} and estimate the initial velocity at each iteration, we employ the Nelder–Mead numerical method \citep{Gao2012} as implemented in \texttt{SciPy}.


\section{Statistical comparison of sub-populations}
\label{sec:stats}

To assess which physical and observational parameters control the agreement between the $\alpha$–$\beta$ dynamical solutions and the photometrically derived masses and classifications, we quantify the effect of individual scalar variables using the Vargha–Delaney effect size $A_{12}$ \citep{VarghaDelaney2000}. The variables considered include $V_{\infty}-V_{\mathrm{ter}}$, $h_{\mathrm{ter}}$, $h_{\mathrm{bri}}$, $h_{\mathrm{beg}}$, $V_{\mathrm{ter}}$, $V_{\mathrm{bri}}$, $V_{\mathrm{beg}}$, $M_{\mathrm{beg}}$, $M_{\mathrm{ter}}$, duration, length, energy, $Pf$, $PE$, $\gamma$, azimuth, as well as the inferred parameters $\alpha$, $\beta$, $\rho_m$, and $\mu$. For each scalar variable $x$ and for each comparison---(i) converged versus non-converged three-point fits, and (ii) events whose inferred density is consistent with their original $PE$ type versus those that are not---we consider two groups,

\[
G_1 = \{x_i\}_{i=1}^{n_1}, \qquad
G_2 = \{y_j\}_{j=1}^{n_2},
\]
and define
\begin{equation}
A_{12}
=
P(X>Y) + \frac{1}{2}P(X=Y),
\label{eq:A12_def_prob}
\end{equation}
where $X$ and $Y$ are random variables obtained by sampling uniformly from $G_1$ and $G_2$, respectively. Thus, $A_{12}$ measures the probability that a randomly chosen value from $G_1$ exceeds a randomly chosen value from $G_2$ (with ties counted as one half). In practice, $A_{12}$ is estimated from the pooled ranks of the two samples. Let $R_1$ be the sum of ranks of the $n_1$ observations in $G_1$ when all $n_1+n_2$ values are ranked together; then
\begin{equation}
\hat{A}_{12}
=
\frac{R_1}{n_1 n_2} - \frac{n_1+1}{2n_2}.
\label{eq:A12_rank}
\end{equation}

We define “non-converged” events as those for which no solution satisfying the convergence criteria defined in Section~\ref{sec:inferring} is obtained.

The Vargha-Delaney $A_{12}$ coefficient is a non-parametric, scale-invariant effect size: it does not assume any specific distribution, and it is invariant under monotonic transformations of $x$. Values $A_{12}=0.5$ indicate no systematic difference between the two groups, while $A_{12}>0.5$ implies that $G_1$ tends to have larger values than $G_2$, and $A_{12}<0.5$ implies the opposite. This makes $A_{12}$ particularly suitable for comparing heterogeneous variables (velocities, heights, masses, orbital elements) and for focusing on effect magnitude rather than on $p$-values, which are strongly influenced by sample size.

The $A_{12}$ measure is directly related to Cliff's delta $\delta$ \citep{Cliff1993} through
\begin{equation}
\delta
=
P(X>Y) - P(X<Y)
=
2A_{12} - 1,
\qquad
A_{12} = \frac{\delta + 1}{2}.
\label{eq:A12_cliff}
\end{equation}

Although the effect size is expressed in terms of $A_{12}$ following \citet{VarghaDelaney2000}, we adopt the magnitude thresholds proposed by \citet{Romano2006}, which are defined for Cliff’s $\delta$. This choice is justified by the direct relationship between both measures, $\delta = 2A_{12} - 1$, which allows a consistent translation of thresholds between the two formulations. The Romano et al. thresholds are widely used in empirical studies and provide a more conservative and standardized interpretation of effect magnitudes. Accordingly, we express the classification in terms of $A_{12}$ by mapping these thresholds into the corresponding probability space.

We therefore adopt the conventional thresholds for $|\delta|$ proposed in the literature \citep{Romano2006} to classify the magnitude of the effect and translate them into the corresponding ranges of $A_{12}$. The categories are:
\begin{align}
|\delta| < 0.147 
&\quad\Leftrightarrow\quad |A_{12} - 0.5| < 0.0735
&&\text{negligible}, \label{eq:thr_negl}\\
0.147 \le |\delta| < 0.33 
&\quad\Leftrightarrow\quad 0.0735 \le |A_{12} - 0.5| < 0.165
&&\text{small}, \label{eq:thr_small}\\
0.33 \le |\delta| < 0.474 
&\quad\Leftrightarrow\quad 0.165 \le |A_{12} - 0.5| < 0.237
&&\text{medium}, \label{eq:thr_medium}\\
|\delta| \ge 0.474 
&\quad\Leftrightarrow\quad |A_{12} - 0.5| \ge 0.237
&&\text{large}. \label{eq:thr_large}
\end{align}

In the following, we report $A_{12}$ for each variable and classify its magnitude according to Eqs.~(\ref{eq:thr_negl})–(\ref{eq:thr_large}). This allows us to identify which parameters show negligible differences between groups and which exhibit small, medium, or large systematic shifts, while preserving a consistent, distribution-free interpretation across all variables.

\section{Results and discussion} \label{sec:results}

The three-point inversion applied to the EN catalog yields high convergence rates when only height and velocity are fitted: 88\% of events satisfy all constraints (percentages are computed with respect to the filtered dataset of 809 events). When masses are included as additional observables, the feasible region of parameter space contracts and the convergence fraction decreases to 63\%. Among the successfully fitted events, 52\% of the inferred bulk densities fall within the ranges consistent with their original $PE$ classifications (34\% of the filtered dataset). 

To quantify the impact of including mass constraints on the inferred $\alpha$–$\beta$ solutions, we computed the relative differences between the parameters obtained with and without mass enforcement. The distributions are centered at zero for both parameters, with median variations consistent with numerical precision and mean shifts below 1\% ($-0.32\%$ for $\alpha$ and $+0.52\%$ for $\beta$). The dispersion remains limited, with standard deviations of 3.5\% and 8.4\% for $\alpha$ and $\beta$, respectively. These results indicate that the inclusion of mass constraints does not introduce a systematic bias in the inferred parameters, and that both inversion approaches yield statistically consistent $\alpha$–$\beta$ solutions, with differences confined to a small subset of events.

\begin{figure}[!t]\centering
  \includegraphics[,width=1\linewidth]{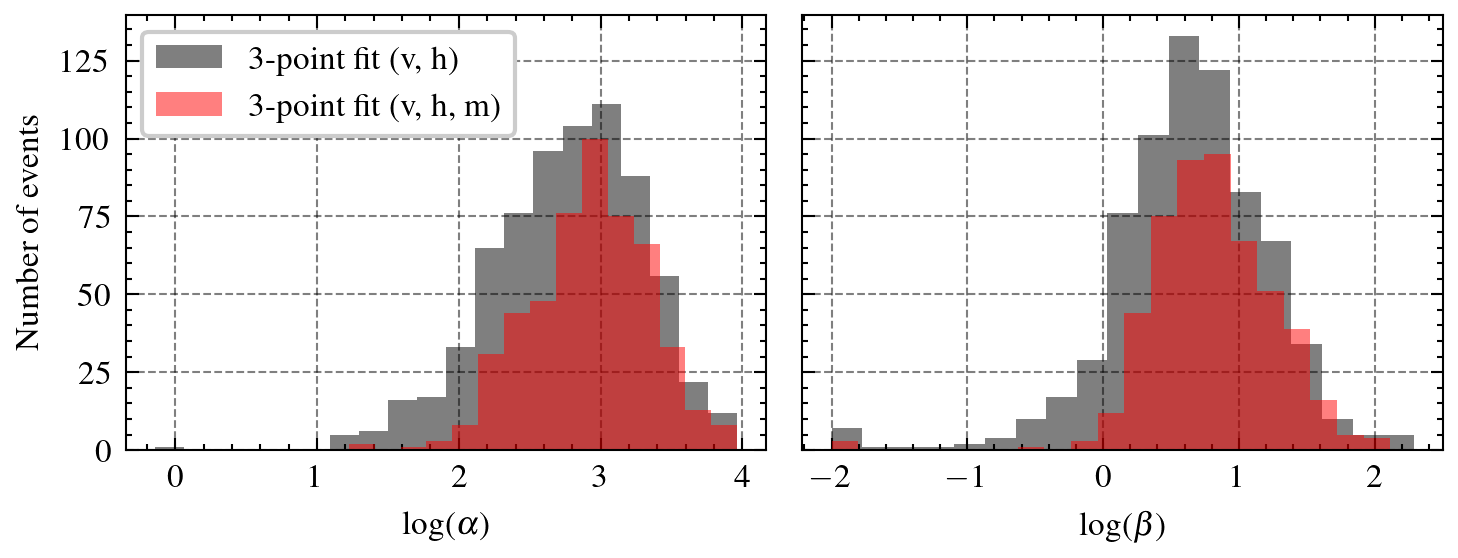}
  \caption{Comparison of the $\alpha$–$\beta$ distributions obtained from the 3-point inversion with and without mass constraints. The left panel shows the distribution of $\log(\alpha)$ and the right panel $\log(\beta)$. The purely kinematic fit (v,h) exhibits a broader distribution with a more extended low-value tail, while the mass-constrained fit (v,h,m) yields a slightly narrower distribution.}
  \label{fig:alpha_beta_compare}
\end{figure}

This behaviour is further illustrated in Fig.~\ref{fig:alpha_beta_compare}, where the distributions obtained with and without mass constraints are directly compared. While both distributions largely overlap and share similar central tendencies, the purely kinematic solutions exhibit a more extended low-value tail, particularly for $\alpha$. The inclusion of mass constraints reduces the occurrence of these low-value solutions, leading to a slightly narrower distribution. This confirms that the primary effect of enforcing mass consistency is to constrain the admissible parameter space rather than to shift the inferred values.

\begin{figure}[!t]\centering
  \includegraphics[,width=1\linewidth]{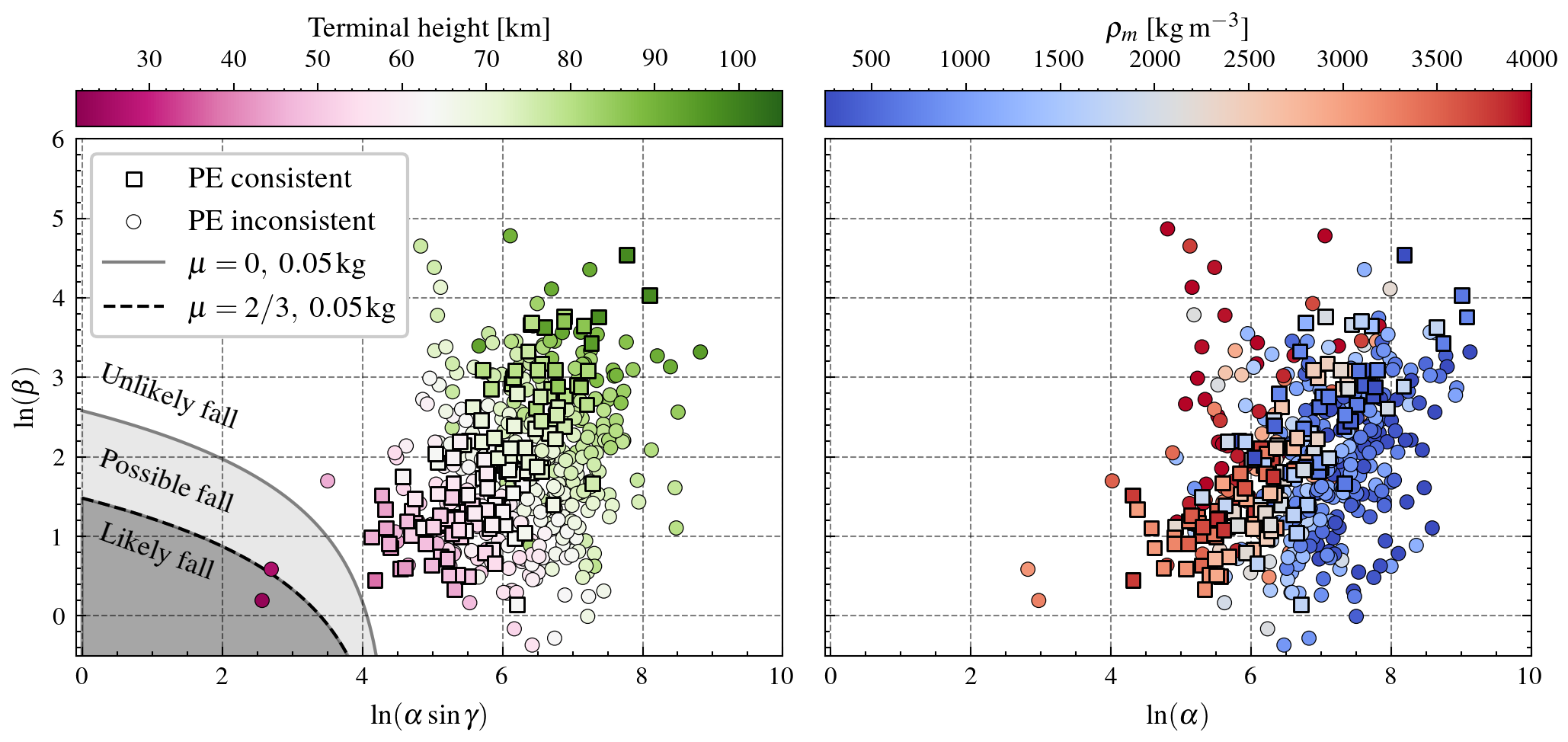}
  \caption{Fitted $\alpha$–$\beta$ distributions from the 3-point fit with mass. The left panel removes trajectory slope dependence. Points are color-coded by terminal height (left) and bulk density (right). Thresholds correspond to a 0.05 kg, 3500 kg\,m$^{-3}$ meteoroid. }
  \label{fig:alpha_beta}
\end{figure}

Figure~\ref{fig:alpha_beta} shows the $\alpha$–$\beta$ distributions obtained from the three-point inversion when mass constraints are included. The left panel presents $\ln(\beta)$ as a function of $\ln(\alpha\sin\gamma)$, which removes the explicit dependence on trajectory slope and allows a direct comparison with the analytical fall–region boundaries traditionally used in the $\alpha$–$\beta$ framework. Points are color–coded by terminal height, providing a proxy for atmospheric penetration depth. Fireballs terminating at high altitudes occupy the upper region of the diagram, consistent with weak ablation (low $\beta$) and modest dynamical loading, whereas events that reach lower terminal heights require larger $\beta$ values, indicating sustained ablation and progressively higher mass–loss rates. Most events cluster above the survival thresholds for a 0.05\,kg fragment at both $\mu = 0$ and $\mu = 2/3$, implying that only a small subset of the catalog is compatible with meteorite–dropping outcomes under standard physical assumptions.

The right panel displays $\ln(\beta)$ versus $\ln(\alpha)$, with points color–coded by the inferred bulk density $\rho_m$. A continuous trend is observed: higher densities generally coincide with larger $\alpha$ values and moderate to high $\beta$ values, indicating that dense meteoroids experience relatively strong drag scaling (through their mass–to–area ratio and entry geometry) while retaining enough structural strength to undergo substantial ablation before full deceleration. Conversely, low–density objects concentrate in the low–$\alpha$, low–$\beta$ region, where early fragmentation, rapid ablation, and high–altitude termination dominate their dynamical evolution. The distribution remains continuous rather than discrete, showing that bulk–density variations alone do not produce isolated dynamical subpopulations within the EN dataset. Instead, density correlates smoothly with the combined aerodynamic and ablative response captured by $\alpha$ and $\beta$. Objects classified as $PE$–consistent (squares) tend to populate the central region of the cloud, reflecting internal coherence between dynamical and photometric constraints and indicating that their kinematic evolution is well captured by the inferred $\alpha$ and $\beta$ values, although a slight shift toward lower $\alpha$ is also noticeable in both panels.

It is important to emphasize the physical meaning of low versus high $\alpha$, as this parameter is often misinterpreted. Although one might expect large $\alpha$ values to correspond to deeper atmospheric penetration, the definition of $\alpha$ (Eq.~\ref{a}) shows that it scales with the drag–to–mass ratio, approximately $\alpha \propto S/M \cdot 1/\sin\gamma$. Thus, large $\alpha$ corresponds to objects with large cross–sectional area relative to mass, or to shallow trajectories, both of which favor early deceleration and high–altitude extinction. Meteorite–producing fireballs therefore require \emph{low} $\alpha$, reflecting compact geometry, high mass–to–area ratio, and steep entry; combined with low $\beta$ (weak ablation), these conditions are necessary for sufficient atmospheric penetration to allow fragment survival at the ground. The placement of known meteorite–dropping analogues in the low–$\alpha$, low–$\beta$ region of the $\alpha$–$\beta$ diagram is consistent with this interpretation.

Taken together, both projections in Figure~\ref{fig:alpha_beta} demonstrate that the inferred $\alpha$ and $\beta$ values follow physically consistent trends: terminal height scales primarily with $\beta$, bulk density correlates with $\alpha$, and the joint distribution reflects the expected coupling between drag, ablation, and material properties. The systematic concentration of $PE$–consistent events near the center of the cloud further supports the internal coherence of the method, indicating that cases where dynamical and photometric mass estimates agree tend to occupy stable dynamical regimes in the $\alpha$–$\beta$ space.


\begin{figure}[!t]\centering
  \includegraphics[,width=0.75\linewidth]{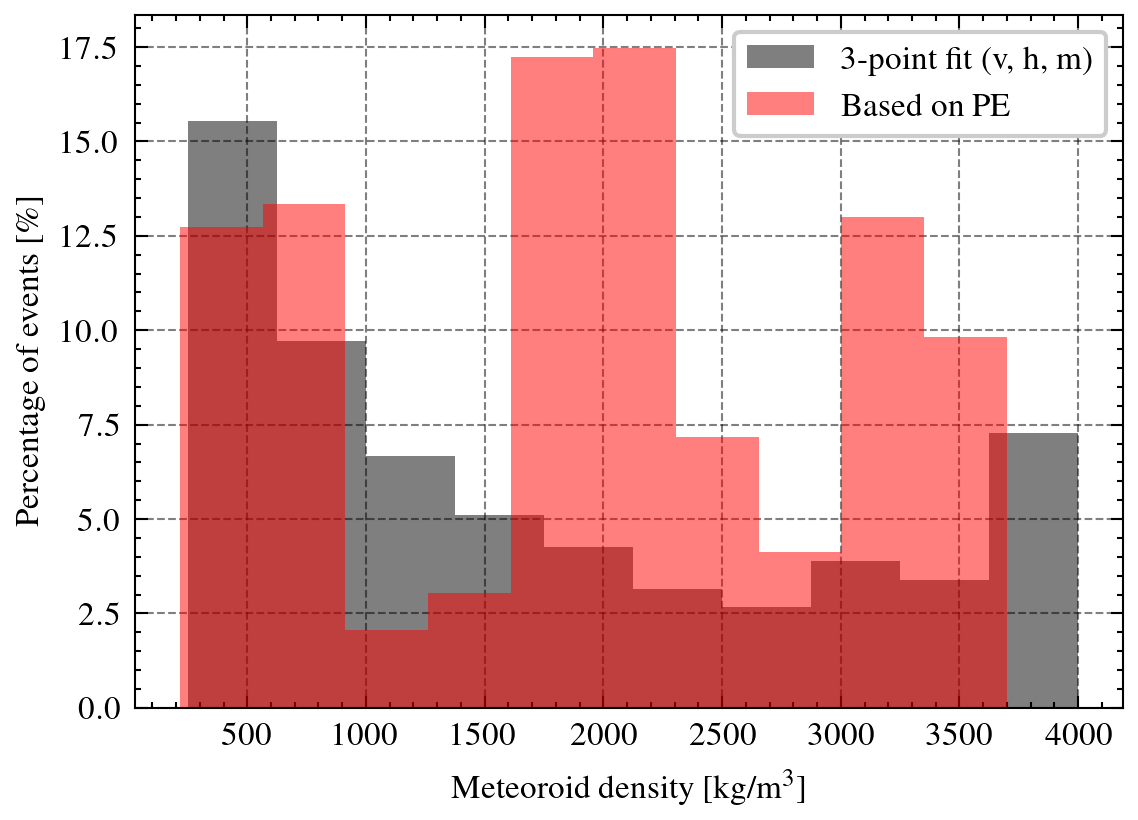}
  \caption{Distribution of meteoroid bulk densities estimated from 3-point fits with mass compared to values derived from $PE$ classification. }
  \label{fig:density}
\end{figure}

Figure~\ref{fig:density} compares the bulk–density distribution inferred from the three-point inversion with mass to the discrete density classes assigned via the $PE$ criterion. The $PE$–based histogram exhibits a set of sharp, artificial peaks at the canonical densities associated with each fireball type: $\sim 3700$~kg\,m$^{-3}$ for type~I, $\sim 2000$~kg\,m$^{-3}$ for type~II, $\sim 750$~kg\,m$^{-3}$ for type~IIIA, and $\sim 270$~kg\,m$^{-3}$ for type~IIIB, with the intermediate categories (I/II and IIIA/IIIB) spanning interpolated ranges between these fixed values \citep{Ceplecha1998SSRv}. Because each $PE$ class maps directly onto one of these density levels, the resulting distribution is strongly discretised and reflects the imposed classification boundaries rather than any measured physical variability.

The three-point dynamical inversion, in contrast, produces a continuous distribution ranging from $\sim 300$ to $\sim 4000$~kg\,m$^{-3}$. Two broad components emerge: a low–density population ($\lesssim 1000$~kg\,m$^{-3}$) and a dense concentration near ordinary–chondrite values ($\gtrsim 3500~kg\,m^{-3}$). Intermediate densities are present but do not form a dominant peak, reflecting a smoother physical transition between weak and strong materials, in agreement with previous studies \citep{Kikwaya2011, Moorhead2017MNRAS4723833M}.

A direct comparison with our previous two-point inversion results \citep{EloyMaria2025} highlights the impact of degeneracy in the earlier formulation. In that work, the inferred bulk densities corresponded to ranges of compatible solutions, and the values reported were representative averages of converged regions in parameter space rather than unique estimates. By contrast, the present three-point inversion yields event-specific density values. Quantitatively, we find that 37\% of events produce compatible density estimates between the two-point and three-point approaches, while only 23\% remain simultaneously consistent with both inversion methods and their original $PE$ classification. This partial overlap indicates that a subset of fireballs is robustly characterized across methodologies, whereas for a significant fraction the additional peak-brightness constraint reduces degeneracy and shifts the solution toward a more restricted and physically consistent parameter space. In addition, in the absence of external constraints in the two-point formulation, the optimization tended to favor high-density solutions.

Both inversion schemes rely on reported observational inputs that often lack homogeneous or explicitly quantified uncertainties. Consequently, the derived parameters should be interpreted with caution, as unknown or underestimated errors in velocity, height, mass, or brightness can propagate nonlinearly into the inferred densities. A natural extension is a fully probabilistic inversion framework in which measurement uncertainties, when available, are explicitly incorporated and propagated into posterior confidence intervals for all retrieved parameters.

Previous observational and modelling studies indicate that meteoroid bulk densities span a wide and continuous range, reflecting the diversity of materials and parent-body origins in the Solar System. For cometary and asteroidal populations, densities typically extend from a few hundred kg\,m$^{-3}$ for porous cometary material up to $\sim 3000$–$4000$~kg\,m$^{-3}$ for compact, chondritic bodies. For example, detailed modelling of faint meteors combining light curves and deceleration measurements yields densities within this range, with clear differences between cometary and asteroidal populations \citep{Kikwaya2011}. Including metallic meteoroids, which are less frequent but significantly denser, the upper bound of the distribution can extend to $\sim 7000$–$8000$~kg\,m$^{-3}$. This naturally leads to a broad, and in some cases bimodal, distribution, although the exact shape depends on the observational sample and the adopted ablation model. At the same time, high-resolution studies have shown that fragmentation processes are complex and not fully captured by current models, which introduces significant uncertainty in density retrievals and can broaden the inferred distributions \citep{Campbell2013AA557A41C}. Overall, these results support the interpretation of meteoroid density as a continuous variable with substantial dispersion rather than as a set of discrete classes, with two main dominant density populations, in agreement with our findings.

\begin{figure}[!t]\centering
  \includegraphics[,width=0.75\linewidth]{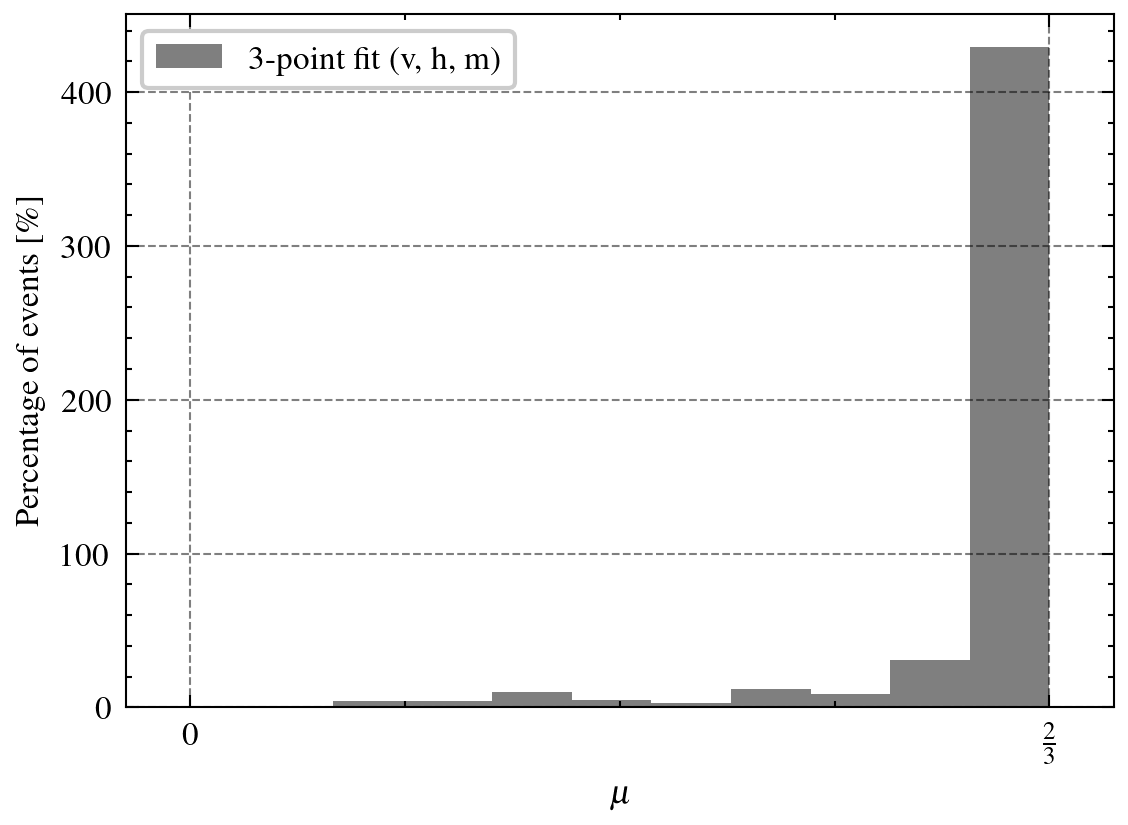}
  \caption{Distribution of shape change coefficients estimated from 3-point fits with mass. }
  \label{fig:mu}
\end{figure}

Figure \ref{fig:mu} shows the distribution of the shape–change coefficient $\mu$ inferred from the three-point inversion with mass. The solutions cluster overwhelmingly near the upper physical limit $\mu = 2/3$, indicating that most meteoroids behave as rapidly rotating bodies whose surface ablates quasi-isotropically. This outcome is fully consistent with previous analyses of well-observed fireballs: \citet{Bouquet2014PSS} quantified $\mu$ for a large Canadian Network dataset and found average values near 0.56, with roughly two-thirds of events converging to $\mu \approx 0.65$, the rotationally dominated limit. The small fraction of solutions at intermediate or low $\mu$ in our sample likely reflects events with limited rotational surface renewal or early fragmentation, where ablation becomes less uniform. The strong preference of the inversion for $\mu \rightarrow 2/3$ therefore supports the physical plausibility of the reconstructed trajectories, and indicates that the EN fireball population shares the same dominant rotational–ablation regime documented by other networks.

Among the seven EN events originally classified as iron-type according to their $PE$ class, only two---EN180318\_001818 and EN121118\_185325---yield densities from the (v,h,m) inversion that are compatible with metallic meteoroids. Even in these cases, EN180318\_001818 also admits solutions within the stony, non-metallic regime. Two additional events, EN171017\_205152 and EN141118\_195214, do not retain iron-like densities but do converge to physically plausible solutions corresponding to non-metallic impactors. The remaining iron-labelled fireballs fail to converge in the (v,h,m) inversion. 

\begin{figure}[!t]\centering
  \includegraphics[,width=0.42\linewidth]{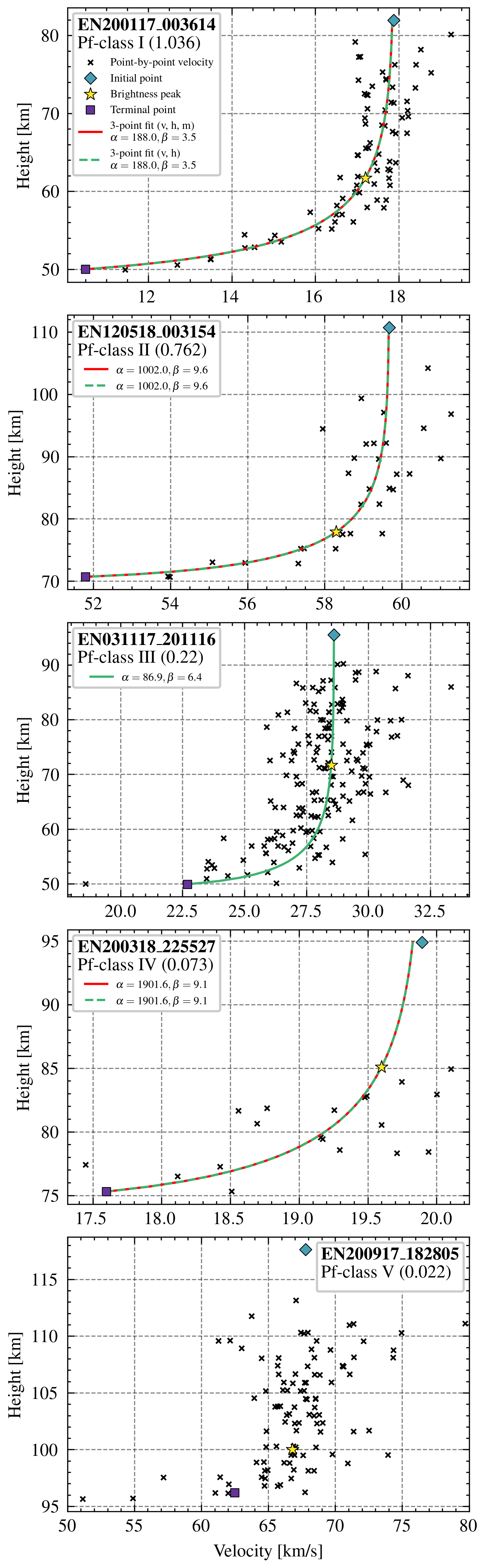}
  \caption{Reconstructed velocity profiles for five events with different $Pf$ classes. Three-point fits with and without mass are displayed. }
  \label{fig:5_examples}
\end{figure}

Figure~\ref{fig:5_examples} illustrates the behaviour of the inversion for five representative EN fireballs spanning a wide range of mechanical strengths, ordered from top to bottom by decreasing $Pf$ class (from~I to~V). For each event, the available point-by-point velocity measurements---kindly provided as height--velocity pairs by Jiří Borovička---are shown as black crosses, but only the three reference points (initial, peak-brightness, and terminal) are used in the inversion itself. The coloured curves correspond to the reconstructed trajectories obtained with the three-point method, with (red solid line) and without (green dashed line) imposing the catalog photometric mass as an additional constraint.

In the first, second, and fourth examples (Pf classes~I, II, and IV), both inversions converge to nearly identical $(\alpha,\beta)$ solutions, indicating that the dynamical constraints supplied by the three points alone already define a narrow admissible region of parameter space. This agreement also shows that, for mechanically coherent or moderately strong bodies, the photometric mass does not conflict with the deceleration implied by the point-by-point data. The reconstructed profiles follow the distribution of measured velocities closely, despite only using three of the points to compute the fit.

The Pf-class~III case behaves differently. Here the inversion without mass constraint succeeds and reproduces the curvature of the measured velocity profile, but no solution satisfying the residual thresholds can be found when the photometric mass is imposed. This mismatch indicates that the catalog mass is incompatible with the kinematic evolution of the event based on $\alpha$-$\beta$, likely due to strong fragmentation or poorly constrained light-curve energetics, as the first part of the event is missing. 

The Pf-class~V example (bottom panel) does not converge under any configuration. Despite a well-populated set of point-by-point velocity measurements, no exponential mass-loss solution can accommodate the strong curvature, rapid ablation, and structural weakening implied by the data. This behaviour is fully consistent with its very low $Pf$ value: mechanically weak objects may undergo complex fragmentation sequences that produce velocity profiles lying far from the analytic trajectory family parameterised by $\alpha$-$\beta$. Moreover, the substantial scatter in the observational data implies correspondingly large uncertainties in the retrieved parameters \citep{Eloy2025RMxAC5987P}. In such cases, the algorithm becomes especially sensitive to how the initial velocity is defined, since small differences in the adopted entry speed propagate through the inversion and can significantly affect convergence behaviour and the resulting physical estimates.

To quantify systematic differences between these sub–populations, we computed Vargha–Delaney effect sizes $A_{12}$ for all scalar parameters. Tables~\ref{tab:var_comp_vh}–\ref{tab:var_comp_conv_PE} summarize the impact obtained for the three comparisons. In the height–velocity inversion (Table~\ref{tab:var_comp_vh}), the strongest discriminator between converged and non-converged fits is the total deceleration, $\Delta V = V_{\infty} - V_{\mathrm{ter}}$, which shows a large effect. Converged solutions are strongly associated with larger measurable deceleration: an event drawn at random from the converged set has an 84.2\% probability of exhibiting a greater velocity change than one drawn from the non-converged set. This behaviour reflects the fact that the kinematic constraints provide substantially greater leverage when the velocity profile shows a clear decrease; in nearly ballistic cases, where $V_{\infty}\approx V_{\mathrm{ter}}$, the inversion lacks sensitivity and the fit often fails. Height-related variables ($h_{\mathrm{bri}}$, $h_{\mathrm{ter}}$) display medium effects, all with $A_{12}<0.5$, indicating that converged events tend to terminate at lower altitudes than non-converged events. This pattern is consistent with deeper atmospheric penetration producing larger dynamic pressures and more pronounced deceleration. Conversely, non-converged events preferentially terminate high in the atmosphere, where weak deceleration, early fragmentation, or strong ablation lead to height–velocity pairs that the analytic model cannot reconcile.

Duration and $Pf$ also occupy medium positions. Converged events tend to have longer luminous durations, which provide a broader kinematic baseline for constraining the height–velocity relation. Likewise, converged solutions are biased toward higher $Pf$ values, indicating mechanically stronger meteoroids whose trajectories exhibit more sustained deceleration. This trend is consistent with the effect observed for the velocity variables: stronger bodies typically enter with lower geocentric speeds (as objects originating from the main belt generally have lower eccentricities and therefore smaller relative velocities) and retain structural coherence over longer atmospheric paths, producing velocity profiles that remain compatible with the assumptions of the $\alpha$-$\beta$ model.

\begin{table}[!t]
\centering
\begin{tabular}{lcc}
\hline
    Variable & $A_{12}$ & Effect \\
\hline    
    $V_{\infty}-V_{\mathrm{ter}}$ & 0.842 & large \\
    $h_{\mathrm{ter}}$ & 0.265 & medium \\
    $Pf$ & 0.732 & medium \\
    $h_{\mathrm{bri}}$ & 0.276 & medium \\
    Duration & 0.723 & medium \\
    $V_{\mathrm{ter}}$ & 0.291 & medium \\
    $PE$ & 0.694 & medium \\
    $V_{\mathrm{bri}}$ & 0.325 & medium \\
    $h_{\mathrm{beg}}$ & 0.326 & medium \\
    $V_{\mathrm{beg}}$ & 0.340 & small \\
    Azimuth & 0.585 & small \\
    $M_{\mathrm{beg}}$ & 0.573 & negligible \\
    Length & 0.562 & negligible \\
    Energy & 0.442 & negligible \\
    $\gamma$ & 0.485 & negligible \\
    $M_{\mathrm{ter}}$ & 0.493 & negligible \\
\hline
\end{tabular}
\caption{Vargha-Delaney effect sizes $A_{12}$ for the comparison between converged and non-converged three-point fits using only height and velocity (see Sec.~\ref{sec:stats}). Effect-size magnitudes follow the thresholds derived from their correspondence with Cliff’s $\delta$.}
\label{tab:var_comp_vh}
\end{table}

In the mass–constrained inversion (Table~\ref{tab:var_comp_mvh}), the parameters with the largest effect sizes are the radiated energy and the initial mass. Both energy and initial mass yield $A_{12}\approx 0.21-0.22$, indicating that converged solutions correspond to meteoroids with systematically lower energies and smaller photometric masses than those in the non–converged group. This reflects the difficulty of fitting high–energy or high–mass fireballs with an exponential mass–loss law: such events typically undergo substantial fragmentation, producing mass–loss histories that cannot be reconciled with the analytic $\alpha$-$\beta$ approach. Variables related to the height of beginning, maximum brightness, and termination exhibit small effects, with converged solutions preferentially associated with higher-altitude trajectories and higher velocities.

\begin{table}[!t]
\centering
\begin{tabular}{lcc}
\hline
    Variable & $A_{12}$ & Effect \\
\hline
    Energy & 0.214 & large \\
    $M_{\mathrm{beg}}$ & 0.220 & large \\
    $h_{\mathrm{bri}}$ & 0.633 & small \\
    $h_{\mathrm{ter}}$ & 0.630 & small \\
    Duration & 0.384 & small \\
    Length & 0.386 & small \\
    $V_{\mathrm{ter}}$ & 0.592 & small \\
    $V_{\mathrm{bri}}$ & 0.592 & small \\
    $V_{\mathrm{beg}}$ & 0.590 & small \\
    $h_{\mathrm{beg}}$ & 0.586 & small \\
    $M_{\mathrm{ter}}$ & 0.434 & negligible \\
    $V_{\infty}-V_{\mathrm{ter}}$ & 0.553 & negligible \\
    Azimuth & 0.460 & negligible \\
    $Pf$ & 0.465 & negligible \\
    $\gamma$ & 0.486 & negligible \\
    $PE$ & 0.501 & negligible \\
\hline
\end{tabular}
\caption{Vargha-Delaney effect sizes $A_{12}$ for the comparison between converged and non-converged three-point fits using height, velocity, and mass (see Sec.~\ref{sec:stats}). Effect categories follow the Cliff’s $\delta$ thresholds.}
\label{tab:var_comp_mvh}
\end{table}

The comparison between events whose inferred bulk density is consistent with their original $PE$ classification and those for which the type changes (Table~\ref{tab:var_comp_conv_PE}) is dominated by the effect of the recovered bulk density itself, as expected. The variable $\rho_m$ shows the strongest separation, with $A_{12}=0.674$, indicating that PE-consistent events are systematically associated with larger densities. In addition, the beginning height $h_{\mathrm{beg}}$ exhibits a medium effect ($A_{12}=0.333$), suggesting that PE-consistent events tend to start their luminous trajectories at slightly lower altitudes. Among the small but non-negligible effects, the most informative parameters are the ballistic coefficient $\alpha$ and the pressure factor $Pf$. The $\alpha$ parameter exhibits $A_{12}=0.370$, indicating that PE-consistent meteoroids tend to have lower $\alpha$ values, consistent with stronger bodies given that the ballistic coefficient is inversely related to density. The pressure factor shows a complementary trend, with $A_{12}=0.623$, indicating that PE-consistent events tend to reach higher dynamic pressures before disruption or termination. This behavior is consistent with the expectation that stronger meteoroids withstand more intense aerodynamic loading, penetrate deeper, and exhibit more pronounced deceleration.

\begin{table}[!t]
\centering
\begin{tabular}{lcc}
\hline
    Variable & $A_{12}$ & Effect \\
\hline
    $\rho_{m}$ & 0.674 & medium \\
    $h_{\mathrm{beg}}$ & 0.333 & medium \\
    $\alpha$ & 0.370 & small \\
    Duration & 0.629 & small \\
    $Pf$ & 0.623 & small \\
    $V_{\mathrm{beg}}$ & 0.389 & small \\
    $V_{\mathrm{bri}}$ & 0.398 & small \\
    $V_{\mathrm{ter}}$ & 0.398 & small \\
    $h_{\mathrm{ter}}$ & 0.412 & small \\
    $h_{\mathrm{bri}}$ & 0.419 & small \\
    $h_{\mathrm{ter}}$ & 0.422 & small \\
    $\gamma$ & 0.438 & negligible \\
    $PE$ & 0.556 & negligible \\
    $M_{\mathrm{beg}}$ & 0.554 & negligible \\
    $\mu$ & 0.462 & negligible \\
    Energy & 0.470 & negligible \\
    Length & 0.477 & negligible \\
    $V_{\infty}-V_{\mathrm{ter}}$ & 0.483 & negligible \\
    Azimuth & 0.506 & negligible \\
    $\beta$ & 0.506 & negligible \\
    $M_{\mathrm{ter}}$ & 0.496 & negligible \\
\hline
\end{tabular}
\caption{Vargha-Delaney effect sizes $A_{12}$ for converged three-point (v,h,m) fits, comparing events whose inferred bulk density is consistent with their original $PE$ classification to those that are not (see Sec.~\ref{sec:stats}). Effect categories follow the Cliff’s $\delta$ thresholds.}
\label{tab:var_comp_conv_PE}
\end{table}

Overall, the $A_{12}$ analysis identifies a consistent set of controlling parameters across the different comparisons. Convergence in the purely kinematic inversion is primarily governed by measurable deceleration, confirming that the method requires sufficiently strong velocity gradients to constrain the solution. When mass constraints are included, the dominant limiting factors shift to the radiated energy and initial mass, indicating that abrupt fragmentation-driven events cannot be reproduced within the continuous mass-loss framework. Finally, consistency with $P_E$ classification is mainly controlled by the inferred bulk density and, to a lesser extent, by parameters related to mechanical strength ($Pf$) and drag ($\alpha$). These results demonstrate that the success and physical consistency of the inversion are driven by a combination of observational sensitivity (deceleration), intrinsic meteoroid properties (mass, energy, strength), and model limitations (exponential mass-loss assumption). In addition, the reliability of the inferred parameters critically depends on the quality of the input data, particularly the accuracy of velocity measurements, their associated uncertainties, and the robustness of the trajectory reconstruction \citep{Vovk2025Icar44116698V, Visuri2026arXiv260115805V}.

\section{Conclusions} \label{sec:conclusions}

We have presented a three-point inversion framework that reconstructs meteoroid deceleration and mass-loss histories from sparse observations consisting of the entry, peak-brightness, and terminal points. Building on the $\alpha$–$\beta$ formalism, the method combines an analytical ablation model with a derivative-free global optimizer and a numerical inversion of the height–velocity relation. Applied to the 2017–2018 EN catalog, the approach recovers physically consistent solutions for a large fraction of events, even when only three kinematic points are available, and yields estimates of bulk density and shape-change coefficient.

When only height and velocity at the three points are enforced, the inversion converges for 88\% of fireballs. Under these conditions, the main discriminator between converged and non-converged solutions is the total deceleration: deeper penetration and longer luminous paths are much more likely to be fitted successfully. This confirms that the $\alpha$–$\beta$ model is well suited for trajectories exhibiting clear deceleration, while nearly ballistic events with weak velocity gradients lie near the intrinsic sensitivity limits of the formalism.

Including mass as an additional observable tightens the parameter constraints and reduces the convergence fraction to 63\%. In this regime, convergence is strongly disfavoured for high-energy, high-mass fireballs, whose fragmentation histories depart from the assumptions of a exponential mass-loss law. The statistical analysis confirms that radiated energy and initial mass are the dominant factors driving non-convergence in the mass-constrained fits. Slightly more than half of the converged solutions yield bulk densities compatible with their $PE$ classes. Among the events with coherent $PE$ assignments, the principal discriminator is the inferred bulk density: denser meteoroids are significantly more likely to remain consistent with their photometrically derived classifications.

The inversion also provides a revised distribution of meteoroid bulk densities that departs significantly from the discrete values imposed by the $PE$ classification. Instead of a set of fixed density classes, we obtain a continuous spread spanning roughly 300–4000 kg\,m$^{-3}$. This distribution exhibits two broad peaks: a low-density component associated with weak or cometary materials, and a higher-density peak near 3500 kg\,m$^{-3}$, consistent with high-strength impactors. Intermediate densities form a smooth continuum rather than falling into predefined categories. The shape-change coefficient shows an equally clear trend: most solutions cluster near the upper physical limit, which corresponds to quasi-isotropic ablation produced by rapid rotation, where the entire surface is efficiently renewed during flight.

By construction, the EN catalog is now complemented with self-consistent $\alpha$ and $\beta$ estimates for all events that pass the three-point inversion criteria. This extends our previous dynamical modeling by providing $\alpha$–$\beta$ values not only for cases with complete velocity profiles but also for events with sparse kinematic information. The enhanced dataset not only increases the number of EN fireballs characterized in the $\alpha$–$\beta$ space, but also exposes systematic tensions (already noted in previous studies) between main-mass dynamical fits and photometric mass estimates that must be accounted for in future fireball modeling and classification schemes.


\section*{Acknowledgements}

This work was partially supported by the Italian Space Agency (ASI) within the LUMIO project (ASI-PoliMi agreement n. 2024-6-HH.0). We express gratitude to the Spanish Ministry of Science, Innovation and Universities for supporting the project No PID2023-151905OB-I00  and to the Academy of Finland project no. 325806 (PlanetS), which facilitated the development of the analytical methods presented in this paper. The program of development within Priority-2030 is acknowledged for supporting the research at UrFU. MG thanks Adolfo González Rivera (Alhama Academy) for logistical support. We thank Jiří Borovička for sharing point-by-point measurements for five EN fireballs.

\appendix
\section{Summary of definitions and abbreviations}\label{sec:defs}
\begin{table}[H]
\footnotesize
\begin{tabular}{lll}
$A$	&	 $-$ 	&	Shape factor, a cross sectional area to volume ratio $A=S\left(\frac{\rho_m}{m}\right)^{2/3}$.	\\
$c_d $ 	&	 $-$ 	&	 Drag coefficient.	\\
$c_h $ 	&	 $-$ 	&	 Heat-transfer coefficient.	\\
$\overline{Ei}$&	 $-$ 	&	 Exponential integral, $\overline{Ei}(x)=\int_{-\infty}^{x}\frac{e^{z}}{z}\,dz\,$.\\
$h$ 	&	 $-$ 	&	 Meteoroid height above sea level ($m$).	\\
$h_{beg}$ 	&	 $-$ 	&	 Meteoroid height above sea level at the beginning of the luminous trajectory ($km$).	\\
$h_{bri}$ 	&	 $-$ 	&	 Meteoroid height above sea level at the brightest point of the luminous trajectory ($km$).	\\
$h_{ter}$ 	&	 $-$ 	&	 Meteoroid height above sea level at the terminal point of the luminous trajectory ($km$).	\\
$h_0$ 	&	 $-$ 	&	 Scale height of the homogeneous atmosphere ($h_0=7.16\,km$).	\\
$H^*$ 	&	 $-$ 	&	 Effective destruction enthalpy ($J\, kg^{-1}$).	\\
$m$ 	&	 $-$ 	&	 Normalized meteoroid mass, $m = M/M_{beg}$  (dimensionless).	\\
$M$ 	&	 $-$ 	&	 Meteoroid mass ($kg$).	\\
$M_{beg}$ 	&	 $-$ 	&	 Initial entry mass of meteoroid at the beginning of the luminous trajectory ($kg$).	\\
$M_{ter}$ 	&	 $-$ 	&	 Terminal mass of meteoroid at the end of the luminous trajectory ($kg$).	\\
$S$ 	&	 $-$ 	&	 Cross sectional area of the body ($m^2$).	\\
$s$ 	&	 $-$ 	&	 Normalized cross-section of the body.	\\
$v$ 	&	 $-$ 	&	 Normalized meteoroid velocity, $v = V/V_{beg}$  (dimensionless).	\\
$V$ 	&	 $-$ 	&	 Meteoroid velocity ($km\, s^{-1}$).	\\
$V_{beg}$ 	&	 $-$ 	&	 Initial entry velocity of the meteoroid at the beginning of the luminous trajectory ($km\, s^{-1}$).	\\
$V_{bri}$ 	&	 $-$ 	&	 Velocity of the meteoroid at the brightest point of the luminous trajectory ($km\, s^{-1}$).	\\
$V_{ter}$ 	&	 $-$ 	&	 Terminal velocity of the meteoroid at the end of the luminous trajectory ($km\, s^{-1}$).	\\
$y$ 	&	 $-$ 	&	 Normalized meteoroid height, $y = h/h_0$  (dimensionless).	\\
$y_{ter}$ 	&	 $-$ 	&	 Normalized meteoroid terminal height, $y_{ter} = h_{ter}/h_0$  (dimensionless).	\\
$\alpha$	&	 $-$ 	&	Ballistic coefficient.\\
$\beta$	&	 $-$ 	&	Mass loss parameter.	\\
$\gamma $ 	&	 $-$ 	&	Angle of the meteoroid flight to the horizontal.	\\
$z$ 	&	 $-$ 	&	Angle of the meteoroid flight to the vertical.	\\
$\mu$	&	 $-$ 	&	Shape change coefficient representing the rotation of the meteoroid ($0\leq\mu\leq2/3$).	\\
$\rho$ 	&	 $-$ 	&	Atmospheric density ($kg\, m^{-3}$).	\\
$\rho_{beg}$ 	&	 $-$ 	& Atmospheric density at the beginning of the observed luminous phase ($g\, cm^{-3}$).	\\
$\rho_{sl}$ 	&	 $-$ 	&	Atmospheric density at sea level ($\rho_{sl}=  1.29 kg\, m^{-3}$).	\\
$\rho_m$ 	&	 $-$ 	&	Meteoroid bulk density ($kg\,m^{-3}$).	\\
$\sigma$	&	 $-$ 	&	Ablation coefficient, $\sigma = \cfrac{c_h}{H^*c_d}\quad$ ($s^2\,km^{-2}$).	\\
$\tau$ 	&	 $-$ 	&	Luminous efficiency coefficient.	\\
$\mathfrak{r}$ & $-$ & Residuals \\
$PE$ & $-$ & Criterion for fireball classification proposed by \citet{Ceplecha1976JGR}.\\
$ln(2\alpha\beta)$ & $-$ & Criterion for fireball classification proposed by \citet{moreno2020physically}.\\
$Pf$ & $-$ & Criterion for fireball classification proposed by \citet{Borovicka2022AA_II}.\\
\end{tabular}
\end{table}

\bibliographystyle{elsarticle-harv} 
\bibliography{cas-refs}
\end{document}